\documentclass{svproc}
\usepackage{url}

\usepackage[backend=biber,
            url=false,
            isbn=false,
            doi=false,
            backref=false,
            natbib=true,%
            mincitenames=1,
            maxcitenames=1,
            citestyle=numeric-comp,
            sorting=none,%
            block=none]{biblatex}

\renewcommand{\bibfont}{\small}
\AtBeginBibliography{\setcounter{maxnames}{10}}

\DeclareFieldFormat{labelnumberwidth}{#1\adddot}
\usepackage{graphics}
\usepackage{dsfont}
\usepackage[pdftex]{graphicx}
\usepackage{wrapfig}
\DeclareGraphicsExtensions{.pdf,.png,.jpg}
\usepackage{epsfig}
\usepackage[font={small}]{caption}
\usepackage{subcaption}
\usepackage[rightcaption]{sidecap}
\usepackage{pbox}
\usepackage{siunitx}

\usepackage{bigstrut}
\usepackage{mathtools}
\usepackage{amsmath, amssymb, amscd}
\usepackage{ wasysym } %
\usepackage{mathptmx} %
\usepackage{gensymb} 
\usepackage{nicefrac}       %
\numberwithin{equation}{section} 

\DeclareMathAlphabet{\mathcal}{OMS}{lmsy}{m}{n}
\usepackage{textcomp} %

\usepackage{algorithm} %
\usepackage{algorithmicx, algpseudocode} %

\usepackage{array} %
\usepackage{tabularx}
\usepackage{multirow}
\usepackage{multicol}
\usepackage{booktabs}
\usepackage{tabulary}

\usepackage[utf8]{inputenc}
\usepackage[english]{babel} %
\usepackage{units}
\usepackage{bm}
\usepackage{xspace}
\usepackage{flushend}%
\usepackage{balance} 
\usepackage{csquotes}
\usepackage{makeidx}
\usepackage{blindtext}

\usepackage{enumitem}

\usepackage{ragged2e}
\usepackage{soul} %
\usepackage{subfiles} %

\usepackage{url}
\makeatletter
\g@addto@macro{\UrlBreaks}{\UrlOrds}
\makeatother
\usepackage{color}
\usepackage[usenames,dvipsnames,table,xcdraw]{xcolor}
\usepackage{pgfplots} %
\pgfplotsset{compat=newest}

\usepackage{marginnote}
\usepackage{soul} %
\newcommand{\todo}[1]{}
\renewcommand{\todo}[1]{{\color{red} TODO: {#1}}}

\newcommand{\tocite}[1]{%
\textcolor{red}{[cite:\ifthenelse{\equal{#1}{}}{}{#1}?]}
}

\newcommand{\ignore}[1]{}



\usepackage[pdfborder={0 0 0.5}]{hyperref}
\hypersetup{
    colorlinks=true,
    linkcolor=black,
    citecolor=black,
    filecolor=cyan,
    urlcolor=black
}

\newcommand{\myargmin}[1] {\underset{#1}{\text{argmin }}}

\newcommand{\algname}{Adjustable Boundary Condition LMPC (ABC-LMPC)\xspace}
\newcommand{\algnameonly}{Adjustable Boundary Condition LMPC \xspace}
\newcommand{\algabbr}{ABC-LMPC\xspace}

\usepackage{amsmath}
\usepackage{amssymb}
\usepackage{cleveref}

\begin{document}
\mainmatter              %

\title{ABC-LMPC: Safe Sample-Based Learning MPC for Stochastic Nonlinear Dynamical Systems with Adjustable Boundary Conditions}
\titlerunning{\algnameonly}  %
\author{Brijen Thananjeyan*\inst{1} \and Ashwin Balakrishna*\inst{1} \and
Ugo Rosolia\inst{2} \and \\ Joseph E. Gonzalez\inst{1} \and Aaron Ames\inst{2} \and Ken Goldberg\inst{1} \\ \scriptsize{* equal contribution}}
\authorrunning{Thananjeyan*, Balakrishna* et al.} %
\tocauthor{Brijen Thananjeyan*, Ashwin Balakrishna*,
Ugo Rosolia, Joseph E. Gonzalez, Aaron Ames, Ken Goldberg}

\institute{University of California Berkeley, Berkeley CA 94720, USA,\\
\and
California Institute of Technology, Pasadena CA 91125, USA\\
\email{ \{bthananjeyan, ashwin\_balakrishna\} @berkeley.edu}}

\maketitle              %
\begin{abstract}
Sample-based learning model predictive control (LMPC) strategies have recently attracted attention due to their desirable theoretical properties and good empirical performance on robotic tasks. However, prior analysis of LMPC controllers for stochastic systems has mainly focused on linear systems in the iterative learning control setting. We present a novel LMPC algorithm, \algname, which enables rapid adaptation to novel start and goal configurations and theoretically show that the resulting controller guarantees iterative improvement in expectation for stochastic nonlinear systems. We present results with a practical instantiation of this algorithm and experimentally demonstrate that the resulting controller adapts to a variety of initial and terminal conditions on $3$ stochastic continuous control tasks.
\end{abstract}

\section{Introduction}
\label{sec:introduction}

Model predictive control (MPC) has seen significant success in a variety of robotic tasks~\cite{SAVED, handful-of-trials, pddm}, and there is substantial experimental and theoretical work demonstrating that the resulting closed loop system performs well on challenging tasks in stochastic dynamical systems~\cite{SAVED, converging-supervisor, SampleBasedLMPC, RLMPC}.  In this work, we build on the recently proposed learning model predictive control (LMPC) framework~\cite{LearningMPC, RLMPC, SampleBasedLMPC}. We assume a known stochastic dynamics model and design an iteratively improving MPC-based control strategy by estimating safe sets and value functions from past closed-loop trajectories.

The LMPC framework~\cite{LearningMPC, RLMPC, SampleBasedLMPC} presents a novel class of reference-free control strategies which utilize MPC to iteratively improve upon a suboptimal controller for a goal directed task. LMPC algorithms typically operate in the iterative learning control setting with fixed initial and terminal conditions, and provide robust guarantees on iterative improvement (in terms of task cost) for stochastic linear systems~\cite{RLMPC, SampleBasedLMPC} and deterministic nonlinear systems~\cite{LearningMPC} if the MPC problem can be solved exactly. However, while LMPC-based control strategies exhibit a variety of desirable theoretical properties~\cite{LearningMPC, RLMPC, SampleBasedLMPC} and have been shown to work well on practical problems on physical robotic systems~\cite{MPCRacing, SAVED}, they have two key limitations: (1) guarantees for stochastic systems are limited to linear systems while practical systems are often stochastic and nonlinear and (2) start states and goal sets are typically assumed to be identical in each iteration.

We address both of these challenges. First, we extend the results in~\cite{SampleBasedLMPC} to show iterative improvement guarantees for stochastic nonlinear systems. Second, we present a method to expand the set of feasible start states and goal sets during learning while maintaining these guarantees. Finally, we introduce sampled-based approximations to present a practical algorithm to learn safe policies, which reliably complete tasks with varying boundary conditions while satisfying pre-specified constraints. The contributions of this work are (1) a novel multi-start, multi-goal LMPC algorithm, \algname, which optimizes expected costs subject to robust constraints, with (2) guarantees on expected performance, robust constraint satisfaction, and convergence to the goal for stochastic nonlinear systems, (3) a practical algorithm for expanding the allowed set of initial states and goal sets during learning, and (4) simulated continuous control experiments demonstrating that the learned controller can adapt to novel start states and goal sets while consistently and efficiently completing tasks during learning.
\vspace{-0.1in}
\section{Related Work}
\label{sec:related-work}

\noindent\textbf{Model Predictive Control:} There has been a variety of prior work on learning based strategies for model predictive control in the reinforcement learning~\cite{SAVED, handful-of-trials, pddm} and controls communities~\cite{provable-MPC, GP-MPC, Safe-Exp-RL-MPC, cautious-MPC, unitary-MPC, hewing2019cautious}. Prior work in learning for model predictive control has focused on estimating the following three components used to design MPC policies: $\emph{i})$ a model of the system~\cite{GP-MPC,cautious-MPC,kocijan2004gaussian, MPCRacing, unitary-MPC, handful-of-trials, pddm, SAVED}, $\emph{ii})$ a safe set of states from which the control task can be completed using a known safe policy~\cite{bacic2003general,brunner2013stabilizing, wabersich2018linear,blanchini2003relatively} and $\emph{iii})$ a value function~\cite{SAVED, SampleBasedLMPC, RLMPC, polo}, which for a given safe policy, maps each state of the safe set to the closed-loop cost to complete the task.  
The most closely related works, both by Rosolia et. al.~\cite{SampleBasedLMPC, RLMPC}, introduce the learning MPC framework for iterative learning control in stochastic linear systems. Here, MPC is used to iteratively improve upon a suboptimal demonstration by estimating a safe set and a value function from past closed loop trajectories. Robust guarantees are provided for iterative controller improvement if the MPC problem can be solved exactly. Furthermore,~\citet{SAVED} propose a practical reinforcement learning algorithm using these strategies to learn policies for nonlinear systems. However,~\cite{RLMPC, SAVED} are limited to the iterative learning control setting, and although~\cite{SampleBasedLMPC} presents a strategy for controller domain expansion, the method is limited to linear systems and requires the user to precisely specify an expansion direction. In this work, we build on this framework by (1) extending the theoretical results to prove that under similar assumptions, LMPC based controllers yield iterative improvement in expectation under certain restrictions on the task cost function and (2) providing a practical and general algorithm to adapt to novel start states and goal sets while preserving all theoretical guarantees on controller performance.

\noindent\textbf{Reinforcement Learning:} There has been a variety of work from the reinforcement learning (RL) community on learning policies which generalize across a variety of initial and terminal conditions. Curriculum learning~\cite{rev-curr, backplay, learning-curriculum-RL} has achieved practical success in RL by initially training agents on easier tasks and reusing this experience to accelerate learning of more difficult tasks. \citet{rev-curr, backplay} train policies initialized near a desired goal state, and then iteratively increase the distance to the goal state as learning progresses. While these approaches have achieved practical success on a variety of simulated robotic and navigation tasks, the method used to expand the start state distribution is heuristic-based and requires significant hand-tuning. We build on these ideas by designing an algorithm which expands the start state distribution for an MPC-based policy by reasoning about reachability, similar to~\citet{BARC}. However, \cite{BARC} provides a curriculum for model free RL algorithms and does not provide feasibility or convergence guarantees, while we present an MPC algorithm which expands the set of allowed start states while preserving controller feasibility and convergence guarantees. There is also recent interest in goal-conditioned RL~\cite{imagined-goals, universal-value}. The most relevant prior work in this area is hindsight experience replay~\cite{HER}, which trains a goal-conditioned policy using imagined goals from past failures. This strategy efficiently reuses data to transfer to new goal sets in the absence of dense rewards. We use a similar idea to learn goal-conditioned safe sets to adapt to novel goal sets by reusing data from past trajectories corresponding to goal sets reached in prior iterations. 

\noindent\textbf{Motion Planning:} The domain expansion strategy of the proposed algorithm, \algabbr, bears a clear connection to motion planning in stochastic dynamical systems~\cite{lqg-mp, gaussian-belief}. Exploring ways to use \algabbr to design motion planning algorithms which can efficiently leverage demonstrations is an exciting avenue for future work, since the receding horizon planning strategy could prevent the exponential scaling in complexity with time horizon characteristic of open-loop algorithms~\cite{motion-plan-long-hor}.

\vspace{-0.1in}
\section{Problem Statement}
\label{sec:prob-statement}
In this work, we consider nonlinear, stochastic, time-invariant systems of the form:
\begin{align}
    x_{t+1}& = f(x_{t}, u_t, w_t) \label{dynamics_open}
\end{align}
where $x_{t} \in \mathbb{R}^n$ is the state at time $t$, $u_t \in \mathbb{R}^m$ is the control, $w_t \in \mathbb{R}^k$ is a disturbance input, and $x_{t+1}$ is the next state. The disturbance $w_t$ is sampled i.i.d. from a known distribution over a bounded set $\mathcal{W} \subseteq \mathbb{R}^p$. We denote Cartesian products with exponentiation, e.g. $\mathcal{W}^2 = \mathcal{W}\times\mathcal{W}$. We consider constraints requiring states to belong to the feasible state space $\mathcal{X}\subseteq \mathbb{R}^n$ and controls to belong to $\mathcal{U}\subseteq \mathbb{R}^m$. Let $x^j_t$,  $u^j_t$, and $w^j_t$ be the state, control input, and disturbance realization sampled at time $t$ of iteration $j$ respectively. Let $\pi^j : \mathbb{R}^n \to \mathbb{R}^m$ be the control policy at iteration $j$ that maps states to controls (i.e. $u_t^j = \pi^j(x_t^j)$).

Unlike~\cite{SampleBasedLMPC}, in which the goal of the control design is to solve a robust optimal control problem, we instead consider an expected cost formulation. Thus, instead of optimizing for the worst case noise realization, we consider control policies which optimize the given cost function in expectation over possible noise realizations. To do this, we define the following objective function with the following Bellman equation and cost function $C(\cdot, \cdot)$:
\begin{align}
    \label{eq:bellman-equation}
    J^{\pi^j}(x^j_0) = \underset{w_0^j}{\mathds{E}} \left[ C(x^j_0, \pi^j(x^j_0)) + J^{\pi^j}(f(x^j_0, u^j_0, w_0^j)) \right]
\end{align}
However, we would like to only consider policies that are robustly constraint-satisfying for all timesteps. Thus, the goal of the control design is to solve the following infinite time optimal control problem:
\begin{align}
    \label{eq:control-objective-true}
    \begin{split}
    J^{j, *}_{0 \rightarrow \infty}(x^j_0) = \underset{\pi^j(.)}{\text{min}} \quad & J^{\pi^j}(x^j_0)  \\
    \text{ s.t. } ~  & x^j_{t+1} = f(x^j_t, u^j_t, w^j_t) \\
    & u^j_t = \pi^j(x^j_t) \\
    & x^j_t \in \mathcal{X}, u^j_t \in \mathcal{U} \\
    & \forall w^j_t \in \mathcal{W}, t \in \{0, 1, \hdots \}
    \end{split}
\end{align}
In this paper, we present a strategy to iteratively design a feedback policy $\pi^j(.): \mathcal{F}_{\mathcal{G}}^j \subseteq  \mathcal{X} \rightarrow \mathcal{U}$, where $\mathcal{F}_{\mathcal{G}}^j$ is the domain of $\pi^j$ for goal set $\mathcal{G}$ (and also the set of allowable initial conditions). Conditioned on the goal set $\mathcal{G}$, the controller design provides guarantees for (i) robust satisfaction of state and input constraints, (ii) convergence in probability of the closed-loop system to $\mathcal{G}$, (iii) iterative improvement: for any $x^j_0 = x^l_0$ where $j < l$, expected trajectory cost is non-increasing ($J^{\pi^j}(x_0^j) \geq J^{\pi^{j+1}}(x_0^{j+1})$), and (iv) exploration: the domain of the control policy does not shrink over iterations ($\mathcal{F}_{\mathcal{G}}^j \subseteq \mathcal{F}_{\mathcal{G}}^{j+1}$ for all goal sets $\mathcal{G}$ sampled up to iteration $j$). In Section~\ref{subsec:transfer}, we describe how to transfer to a new goal set $\mathcal{H}$ by reusing data from prior iterations while maintaining the same properties.

We adopt the following definitions and assumptions:

\begin{assumption}
\label{ass:cost}
\textbf{Costs: }We consider costs which are zero within the goal set $\mathcal{G}$ and greater than some $\epsilon > 0$ outside the goal set: $\exists \epsilon > 0 \text{ s.t. } C(x, u) \geq \epsilon \mathds{1}_{\mathcal{G}^C}(x)$ where $\mathds{1}$ is an indicator function and $\mathcal{G}^C$ is the complement of $\mathcal{G}$.
\end{assumption}

\begin{definition}
\textbf{Robust Control Invariant Set: }As in~\citet{SampleBasedLMPC}, we define a robust control invariant set $\mathcal{A} \subseteq \mathcal{X}$ with respect to dynamics
$f(x, u, w)$ and policy class $\Pi$ as a set where $\forall x \in \mathcal{A}, \ \exists \pi \in \Pi \text{ s.t. } f(x, \pi(x), w) \in \mathcal{A}, \pi(x) \in \mathcal{U}, \ \forall w \in \mathcal{W}$.
\end{definition}

\begin{assumption}
\label{ass:robust_control_invariant_goal}
\textbf{Robust Control Invariant Goal Set: }$\mathcal{G}\subseteq\mathcal{X}$ is a robust control invariant set with respect to the dynamics and the set of state feedback policies $\Pi$.
\end{assumption}
\vspace{-0.2 in}

\section{Preliminaries}
\label{sec:preliminaries}
Here we formalize the notion of safe sets, value functions, and how they can be conditioned on specific goals. We also review standard definitions and assumptions.
\vspace{-0.1 in}
\subsection{Safe Set}
\label{subsec:safeset}
We first recall the definition of a robust reachable set as in~\cite{SampleBasedLMPC}:
\begin{definition}
\textbf{Robust Reachable Set: }The robust reachable set $\mathcal{R}^\pi_t(x_0^j)$ contains the set of states reachable in $t$-steps by the system~\eqref{dynamics_open} in closed loop with $\pi$ at iteration $j$:
\begin{align}
\label{robust-reachable-set}
\begin{split}
\mathcal{R}^\pi_{t+1}(x_0^j) = &\left\{ x_{t+1} \in \mathbb{R}^n | \ \exists w_t \in \mathcal{W}, \ x_t \in \mathcal{R}^\pi_t (x_0^j), \ x_{t+1} = f(x_{t}, \pi(x_t), w_t) \right\}
\end{split}
\end{align}
where $\mathcal{R}^\pi_0(x_0^j) = x_0^j$. We define $\mathcal{R}^\pi_{t+1}$ similarly when the input is a set and for time-varying policies.
\end{definition}

Now, we define the safe set at iteration $j$ for the goal set $\mathcal{G}$ as in~\cite{SampleBasedLMPC}.

\begin{definition}
\textbf{Safe Set: }The safe set $\mathcal{SS}^j_{\mathcal{G}}$ contains the full evolution of the system at iteration $j$,
\vspace{-0.2 in}
\begin{align}
\label{eq:safeSetDef}
\mathcal{SS}^j_{\mathcal{G}} = \left\{ \bigcup_{t=0}^{\infty} \mathcal{R}^{\pi^j}_t(x_0^j) \ \bigcup \mathcal{G} \right\}.
\end{align}
\end{definition}
Note that~\eqref{eq:safeSetDef} is robust control invariant by construction~\cite{SampleBasedLMPC}. We could set $\mathcal{SS}_{\mathcal{G}}^0 = \mathcal{G}$ or initialize the algorithm with a nominal controller $\pi^0$. This enables the algorithm to naturally incorporate demonstrations to speed up training.
\begin{definition}
\textbf{Expected Cost: }The expected cost of $\pi^j$ from start state $x^j_0$ is defined as
\begin{align}
J^{\pi^j}(x_0^j) &= \underset{w^j}{\mathbb{E}}\left[\sum_{t=0}^\infty C(x^j_t, \pi^j(x_t^j))\right] \label{policyval}
\end{align}
\end{definition}
\vspace{-0.3 in}
\subsection{Value Function}
\label{subsec:value_func}
\begin{definition}
\textbf{Value Function: }Recursively define the \textit{value function} of $\pi^j$ in closed-loop with~\eqref{eq:control-objective-true} as:
\begin{align}
     \label{def:value-function}
     L_{\mathcal{G}}^{\pi^{j}}(x) = \begin{cases} 
          \underset{w}{\mathds{E}}\left[C(x, \pi^j(x)) + L_{\mathcal{G}}^{\pi^j}(f(x, \pi^j(x), w))\right] & x\in \mathcal{SS}_{\mathcal{G}}^j \\
          +\infty & x \not\in\mathcal{SS}_{\mathcal{G}}^j
       \end{cases}
\end{align}
Let $\textstyle V_{\mathcal{G}}^{\pi^j}(x) = \underset{k\in \{0, \hdots j\}}{\min} L_{\mathcal{G}}^{\pi^k}(x)$, which is the expected cost-to-go of the best performing prior controller at $x$. 
\end{definition}
Observe that $L^{\pi^j}_\mathcal{G}$ is defined only on $\mathcal{SS}^j_\mathcal{G}$, and $J^{\pi^j} = L^{\pi^j}_\mathcal{G}$ on $\mathcal{SS}^j_\mathcal{G}$. In the event a nominal controller $\pi^0$ is used, we require the following assumption on the initial safe set $\mathcal{SS}^0_{\mathcal{G}}$, which is implicitly a restriction on $\pi^0$ for start state $x_0^0$.

\begin{assumption}
\label{ass:safe_set_initial_condition}
\textbf{Safe Set Initial Condition: }If a nominal controller $\pi^0$ is used, then $\forall x \in \mathcal{SS}^0_{\mathcal{G}},\ L_{\mathcal{G}}^{\pi^0}(x) < \infty$.
\end{assumption}
This assumption requires that the nominal controller is able to robustly satisfy constraints and converge in probability to $\mathcal{G}$. If no nominal controller is used, then this assumption is not required. In that case, we let $\mathcal{SS}^0_{\mathcal{G}} = \mathcal{G}$ and $L_{\mathcal{G}}^{\pi^0}(x) = 0\ \forall x \in \mathcal{SS}^0_{\mathcal{G}}$.
\subsection{Transfer to Novel Goal Sets}
\label{subsec:transfer}
While \citet{SampleBasedLMPC} studies tasks with fixed goal sets, here we show how the safe set and value function can be modified to transfer the learned controller at iteration $j+1$ to a new robust control invariant goal set $\mathcal{H}$ and reuse data from the earlier iterations to accelerate learning.

\begin{definition}
\label{def:safe-set-goal-conditioned}
\textbf{Goal Conditioned Safe Set: }Define the goal conditioned safe set by collecting the prefixes of all robust reachable sets until they robustly fall in $\mathcal{H}$ as follows:
\begin{align}
    \mathcal{SS}_{\mathcal{H}}^{j} = \begin{cases}
        \bigcup_{k=0}^{k^*} \mathcal{R}^{\pi^j}_{k} \bigcup \mathcal{H} & \underset{{k \in \mathbb{N}}}{\max}\mathds{1}\{\mathcal{R}_k^{\pi^j}(x_0^j) \subseteq \mathcal{H}\} = 1 \\
        \mathcal{H} & \text{otherwise}
    \end{cases}
\end{align}
where $\textstyle k^* = \underset{k \in \mathbb{N}}{\arg\max}\mathds{1}\{\mathcal{R}_k^{\pi^j}(x_0^j) \subseteq \mathcal{H}\}$
\end{definition}

We also redefine the value function as follows:
\begin{definition}
\label{def:value-function-goal-cond}
\textbf{Goal Conditioned Value Function: }Recursively define the \textit{goal-conditioned value function} of $\pi^j$ in closed-loop with~\eqref{eq:control-objective-true} as:
\begin{align}
     L^{\pi^{j}}_\mathcal{H}(x) = \begin{cases} 
          \underset{w}{\mathds{E}}\left[C(x, \pi^j(x)) + L^{\pi^j}(f(x, \pi^j(x), w))\right] & x\in \mathcal{SS}_{\mathcal{H}}^j \setminus \mathcal{H} \\
          0 & x \in \mathcal{H}\\
          +\infty & x \not\in\mathcal{SS}^j_{\mathcal{H}}
       \end{cases}
\end{align}
Define $V^{\pi^j}_\mathcal{H}(x) = \underset{k\in \{0, \hdots j\}}{\min} L^{\pi^k}_{\mathcal{H}}(x)$ as before.
\end{definition}
This new value function is for a policy that executes $\pi^j$ but switches to a policy that keeps the system in $\mathcal{H}$ upon entry.
\section{Controller Design}
\label{sec:controller_design}
\begin{figure}
  \centering
  \includegraphics[width=0.95\linewidth]{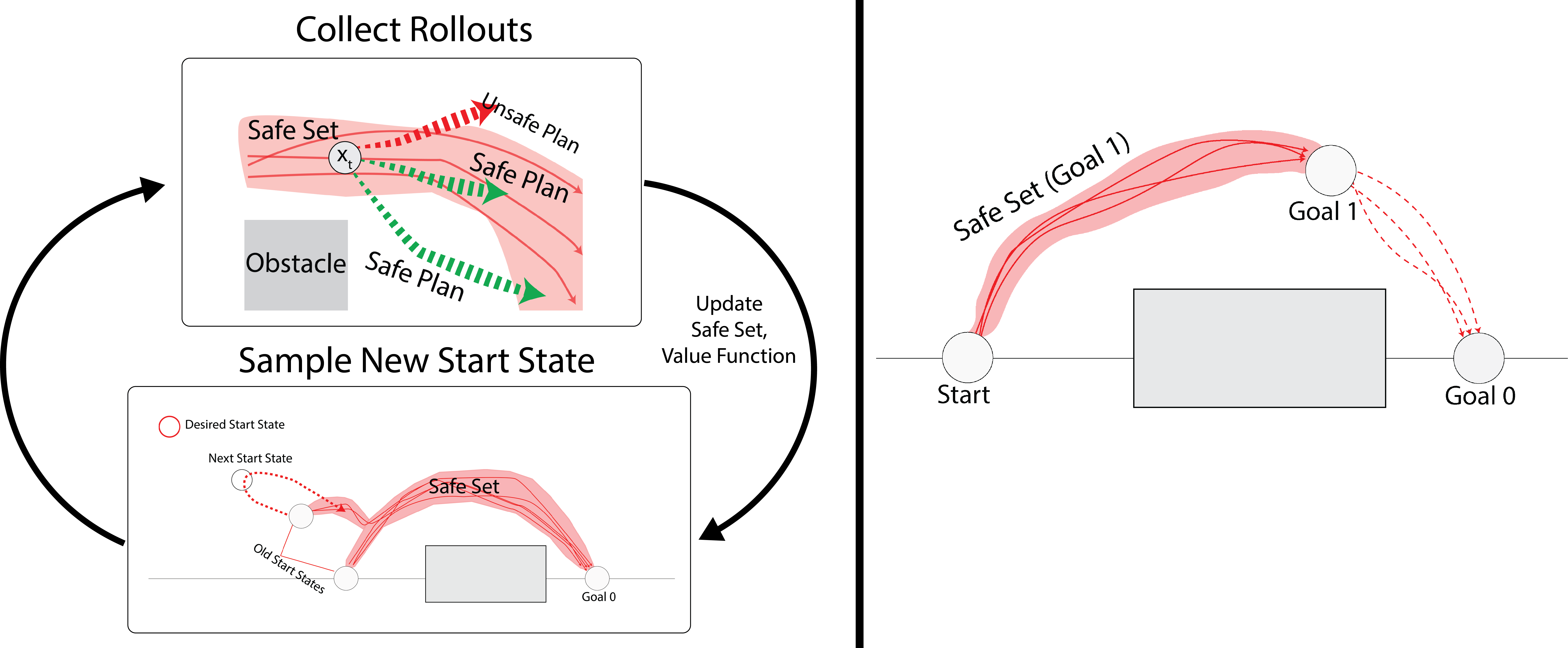}
\caption{\textbf{ABC-LMPC Iterative Algorithm (Left): }ABC-LMPC alternates between (1) collecting rollouts under the current policy $\pi^j$ given $\mathcal{SS}^{j}_{\mathcal{G}_0}$ and $L^{\pi^j}_{\mathcal{G}_0}$ (by optimizing~\eqref{eq:control-objective}), (2) updating $\mathcal{SS}^{j+1}_{\mathcal{G}_0}$ and $L^{\pi^{j+1}}_{\mathcal{G}_0}$ given the new rollouts, and (3) expanding the controller domain towards a desired start state (by optimizing~\eqref{eq:reachability-policy-objective}); \textbf{Goal Set Transfer (Right): }When a new goal set $\mathcal{G}_1$ is supplied, trajectories to goal $\mathcal{G}_0$ can be reused to estimate a new safe set for a new goal $\mathcal{G}_1$ ($\mathcal{SS}^{j}_{\mathcal{G}_1}$) and associated value function ($L^{\pi^j}_{\mathcal{G}_1}$).}
\label{fig:method-splash}
\end{figure}

Here we describe the controller design for optimizing the task cost function while satisfying state and input constraints (Section~\ref{subsec:controller_design_fixed_start}), and discuss how this can be extended to iteratively expand the controller domain (Section~\ref{subsec:expansion}). We consider a fixed goal set $\mathcal{G}$ for clarity, but note that the same formulation holds for other goal sets if the safe set and value function are appropriately defined as in Definitions~\ref{def:safe-set-goal-conditioned}~and~\ref{def:value-function-goal-cond}. See Figure~\ref{fig:method-splash} for an illustration of the full ABC-LMPC controller optimization procedure.

\subsection{Task Driven Optimization}
\label{subsec:controller_design_fixed_start}
At time $t$ of iteration $j$ with goal set $\mathcal{G}$, the controller solves the following receding-horizon trajectory optimization problem with planning horizon $H > 0$:
\begin{align}
    \label{eq:control-objective}
    \begin{split}
    J^{j}_{t\to t + H}(x_t^j) = \min_{\pi_{t:t+H-1|t}\in \Pi^H} \quad &\underset{w^j_{t:t+H-1}}{\mathbb{E}}\left[\sum_{i=0}^{H -1} C(x^j_{t+i|t}, \pi_{t+i|t}(x^j_{t+i|t})) + V_{\mathcal{G}}^{\pi^{j-1}} (x^j_{t+H|t})\right]\\
     \text{s.t. } \quad \quad \  & x^j_{t+i+1|t} = f(x^j_{t+i|t}, \pi_{t+i|t}(x^j_{t+i}), w_{t+i})\ \forall i \in \{0,\ldots,H-1\}\\
    & x^j_{t+H|t} \in \bigcup_{k=0}^{j-1}\mathcal{SS}^{k}_{\mathcal{G}},\ \forall w^j_{t:t+H-1} \in \mathcal{W}^{H}\\
    & x^j_{t:t+H|t} \in \mathcal{X}^{H+1},\ \forall w^j_{t:t+H-1} \in \mathcal{W}^{H}
    \end{split}
\end{align}
where $\pi_{t + i|t}$ is the $i$-th policy in the planning horizon conditioned on $x_t^j$ and $\pi_{t:t+H-1|t} = \{\pi_{t|t},\ldots,\pi_{t+H-1|t}\}$ (likewise for other optimization variables). Let the minimizer of~\eqref{eq:control-objective} be $\pi_{t:t+H-1|t}^{*,j}$. Then, execute the first policy at $x_t^j$:
\begin{equation}\label{eq:MPC_policy}
    u_t^j = \pi^j(x^j_t) = \pi^{*,j}_{t|t}(x^j_t)
\end{equation}
Solving~\ref{eq:control-objective} is typically intractable in practice, so we discuss practical approximations we make to the designed algorithm in Section~\ref{sec:methods}.
\vspace{-0.15 in}

\subsection{Start State Expansion}
\label{subsec:expansion}
We now describe the control strategy for expanding the controller domain.
If there exists a policy $\pi$ for which the $H$-step robust reachable set for the start states sampled at iteration $j$ is contained within the current safe set for goal set $\mathcal{G}$, then we can define the feasible set/domain for the controller at iteration $j$. The domain of $\pi^j$ for $\mathcal{G}$ is computed by collecting the set of all states for which there exists a sequence of policies which robustly keep the system in $\bigcup_{k=0}^{j-1}\mathcal{SS}^k_{\mathcal{G} }$. Precisely, we define the controller domain as follows:
\vspace{-0.1 in}
\begin{align} 
\label{domain-set}
\mathcal{F}_{\mathcal{G}}^j = \{ x \ | \ \exists \pi_{0:H-1} \in \Pi^H \text{ s.t. } \mathcal{R}^{\pi_{0:H-1}}_{H} (x) \subseteq \bigcup_{k=0}^{j-1}\mathcal{SS}^k_{\mathcal{G} }\}
\end{align}
This set defines the states from which the system can robustly plan back to $\bigcup_{k=0}^{j-1}\mathcal{SS}^k_{\mathcal{G}}$. Note that the controller domain is a function of the goal set $\mathcal{G}$.
While any start state sampled from $\mathcal{F}_{\mathcal{G}}^j$ will ensure feasibility and convergence for goal set $\mathcal{G}$ (proven in Section~\ref{sec:controller_props}), we present a method to compute states from $\mathcal{F}^j_\mathcal{G} \setminus \bigcup_{k=0}^{j-1}\mathcal{SS}^k_{\mathcal{G} }$ to expand $\mathcal{F}^{j}_\mathcal{G}$ towards a desired start state, which may not be added to the domain through task-directed exploration. Computing this set is intractable for general nonlinear stochastic systems, so we introduce the following method to approximate this.

At the end of iteration $j$, we sample a start state $x^j_S \in \bigcup_{k=0}^{j} \mathcal{SS}^k_{\mathcal{G}}$ and seek to execute a sequence of $H'$ exploration policies $\pi_{E, 0:H'-1}^j$ which carry the system outside of $\bigcup_{k=0}^{j} \mathcal{SS}^k_{\mathcal{G}}$ and then robustly back into $\bigcup_{k=0}^{j} \mathcal{SS}^k_{\mathcal{G}}$, for all noise realizations $w_{0:H'-2} \in \mathcal{W}^{H'-1}$ where $H' \geq 0$. The sequence of policies $\pi_{E, 0:H'-1}^j$ is computed by solving an $H'$-step optimization problem with a cost function $C_E^j(x, u)$ that encourages exploration outside of $\bigcup_{k=0}^j \mathcal{SS}^k_{\mathcal{G}}$ while enforcing that the controller terminates in some state $x^j_{H'} \in \bigcup_{k=0}^j \mathcal{SS}^k_{\mathcal{G}}$. In Section~\ref{sec:methods}, we discuss some possibilities for $C_E^j(x, u)$, implement one instantiation, and demonstrate that it enables adaptation to novel start states while maintaining controller feasibility. The sequence of controllers can computed by solving the following 1-step trajectory optimization problem:
\vspace{-0.1 in}
\begin{small}
\begin{align}
\begin{split}
 \pi_{E, 0:H'-1}^j = \myargmin{\pi_{0:H'-1}\in \Pi^{H'}} \quad &\underset{w^j_{0:H'-2}}{\mathds{E}}\left[\sum_{i=0}^{H'-1} C_E^j(x^j_{i}, \pi_{i}(x^j_{i}))\right]\\
\text{s.t. } \quad & x^j_{i+1} = f(x^j_{i}, \pi_{i}(x^j_{i}), w_i), \ \forall i \in \{0,\ldots,H'-1\}\\
& x^j_{H'} \in \bigcup_{k=0}^{j} \mathcal{SS}^k_{\mathcal{G}},\ \forall w_{0:H'-2} \in \mathcal{W}^{H'-1}\\
& x^j_{0:H'} \in \mathcal{X}^{H'+1},\ \forall w_{0:H'-2} \in \mathcal{W}^{H'-1}
\end{split}
\label{eq:reachability-policy-objective}
\end{align}
\end{small}

Let $\mathcal{M} =\left(\mathcal{R}^{\pi_{E, 0:H'-1}^j}_i(x_S^j)\right)_{i=0}^{H'}$, the set of all states reachable in $H'$ steps by $\pi_E$ and let $\mathcal{M}_{H} = \left(\mathcal{R}^{\pi_{E, 0:H'-1}^j}_i(x_S^j)\right)_{i=H'-H}^{H'}$. Note that $\forall x \in \mathcal{M}_{H}$, the controller initialized at $x$ can be robustly guided to $\bigcup_{k=0}^{j} \mathcal{SS}^k_{\mathcal{G}}$ in $H$ steps. At iteration $j+1$, feasible start states can be sampled from $\mathcal{M}_{H}$ to guide the policy's domain toward a desired target start state. An MPC policy $\pi^j_E$ could be executed instead to generate these future start states. We could also use the exploration policy to explicitly augment the value function $L^{\pi^j}_\mathcal{G}$ and safe set $\mathcal{SS}^{j}_\mathcal{G}$ and thus $\mathcal{F}^j_\mathcal{G}$ (Appendix~\ref{appendix:controller_spec}). This could be used for general domain expansion instead of directed expansion towards a desired start state.
\vspace{-0.15 in}
\section{Properties of \algabbr}
\label{sec:controller_properties}
In this section, we study the properties of the controller constructed in Section~\ref{sec:controller_design}. For analysis, we will assume a fixed goal set $\mathcal{G}$, but note that if the goal set is changed at some iteration, the same properties still apply to the new goal set $\mathcal{H}$ by the same proofs, because all of the same assumptions hold for $\mathcal{H}$. See Appendix~\ref{ssec:appendix-proofs} for all proofs.
\label{sec:controller_props}
\begin{lemma}
\label{lemma:recursive_feasibility}
\textbf{Recursive Feasibility: }Consider the closed-loop system~\eqref{eq:control-objective}~and~\eqref{eq:MPC_policy}. Let the \textit{safe set} $\mathcal{SS}^j_{\mathcal{G}}$ be defined as in~\eqref{eq:safeSetDef}. If assumptions~\ref{ass:cost}-\ref{ass:safe_set_initial_condition} hold and $x_0^j \in \mathcal{F}_{\mathcal{G}}^j$, then the controller induced by optimizing~\eqref{eq:control-objective}~and~\eqref{eq:MPC_policy} is feasible almost surely for $t\geq 0$ and $j\geq 0$. Equivalently stated, $\underset{w_{0:H-1}^j}{\mathds{E}} [J^j_{t\rightarrow t+H}(x_t^j)] < \infty$, $\forall t, j \geq 0$.
\end{lemma}

\Cref{lemma:recursive_feasibility} shows that the controller is guaranteed to satisfy state-space constraints for all timesteps $t$ in all iterations $j$ given the definitions and assumptions presented above. Equivalently, the expected planning cost of the controller is guaranteed to be finite. The following lemma establishes convergence in probability to the goal set given initialization within the controller domain.

\begin{lemma} 
\label{lemma:convergence}
\textbf{Convergence in Probability: }Consider the closed-loop system defined by \eqref{eq:control-objective}~and~\eqref{eq:MPC_policy}. Let the \textit{sampled safe set} $\mathcal{SS}^j_{\mathcal{G}}$ be defined as in~\eqref{eq:safeSetDef}. Let assumptions~\ref{ass:cost}-\ref{ass:safe_set_initial_condition} hold and $x_0^j \in \mathcal{F}_{\mathcal{G}}^j$. If the closed-loop system converges in probability to $\mathcal{G}$ at iteration $0$, then it converges in probability at all subsequent iterations. Stated precisely, at iteration $j$: $\lim_{t \rightarrow \infty} P(x^j_t \not\in\mathcal{G}) = 0$.
\end{lemma}

\begin{theorem}
\label{theorem:improvement}
\textbf{Iterative Improvement: }Consider system~\eqref{dynamics_open} in closed-loop with \eqref{eq:control-objective} and \eqref{eq:MPC_policy}. Let the \textit{sampled safe set} $\mathcal{SS}^j$ be defined as in~\eqref{eq:safeSetDef}. Let assumptions~\ref{ass:cost}-~\ref{ass:safe_set_initial_condition} hold, then the expected cost-to-go~\eqref{policyval} associated with the control policy~\eqref{eq:MPC_policy} is non-increasing in iterations for a fixed start state. More formally:
\begin{align*}
\forall j \in \mathbb{N}, \ x^j_0 \in \mathcal{F}_{\mathcal{G}}^j, \ x^{j+1}_0 \in \mathcal{F}_{\mathcal{G}}^{j+1} \implies \ J^{\pi^j}(x^j_0) \geq J^{\pi^{j+1}}(x^{j+1}_0)
\end{align*}
Furthermore, $\{J^{\pi^j}(x^j_0)\}_{j=0}^\infty$ is a convergent sequence.
\end{theorem}

\Cref{theorem:improvement} extends prior results~\cite{SampleBasedLMPC}, which guarantee robust iterative improvement for stochastic linear systems with convex costs and convex constraint sets. Here we show iterative improvement in \textit{expectation} for \algabbr for stochastic nonlinear systems with costs as in Assumption~\ref{ass:cost}. The following result implies that the controller domain is non-decreasing.
\begin{lemma}
\label{lemma:expansion}
\textbf{Controller domain expansion: }The domain of $\pi^j$ is an non-decreasing sequence of sets: $\mathcal{F}_{\mathcal{G}}^j\subseteq\mathcal{F}^{j+1}_\mathcal{G}$.
\end{lemma}
\vspace{-0.2 in}

\section{Practical Implementation}
\label{sec:methods}
\algabbr alternates between two phases at each iteration: the first phase performs the task by executing $\pi^j$ and the second phase runs the exploration policy $\pi^j_{E,0:H'-1}$. Only data from $\pi^j$ is added to an approximation of $\mathcal{SS}^j_{\mathcal{G}}$, on which the value function $L^{\pi^j}$ is fit, but in principle, data from $\pi^j_{E,0:H'-1}$ can also be used. Although the task ~\eqref{eq:control-objective} and exploration~\eqref{eq:reachability-policy-objective} objectives are generally intractable, we present a simple algorithm which introduces sampled-based approximations to expand the policy's domain $\mathcal{F}_{\mathcal{G}}^j$ while approximately maintaining theoretical properties in practice. Here, we describe how each component in the controller design is implemented and how optimization is performed. See Appendix~\ref{ssec:appendix-implementation} for further implementation details. 
\vspace{-0.15in}
\subsection{Sampled-Based Safe Set}
\label{subsec:practical-safeset}
\vspace{-0.05in}
In practice, as in~\cite{SampleBasedLMPC}, we approximate the safe set $\mathcal{SS}^j_{\mathcal{G}}$ using samples from the closed loop system defined by~\eqref{eq:control-objective} and~\eqref{eq:MPC_policy}. To do this, we collect $R$ closed-loop trajectories at iteration $j$, each of length $T$ as in \cite{SampleBasedLMPC} where $T$ is the task horizon.

Thus, given the $i$th disturbance realization sequence collected at iteration $j$, given by $\mathbf{w}^j_i = [w^j_{0, i}, \hdots w^j_{T, i}]$, we define the closed loop trajectory associated with this sequence as in~\cite{SampleBasedLMPC}:     $\mathbf{x}^j(\mathbf{w}^j_i) = \left[ x^j_0(\mathbf{w}^j_i), \hdots,  x^j_{T}(\mathbf{w}^j_i) \right]$. As in~\cite{SampleBasedLMPC}, we note that $x_k^j(\mathbf{w}^j_i) \in \mathcal{R}^{\pi^j}_k(x^j_0)$, so $R$ rollouts from the closed-loop system provides a sampled-based approximation to $\mathcal{R}^{\pi^j}_k(x^j_0)$ as follows: $    \tilde{\mathcal{R}}^{\pi^j}_k(x^j_0) = \bigcup_{i=1}^{R} x^j_k(\mathbf{w}^j_i) \subseteq \mathcal{R}^{\pi^j}_k(x^j_0)$. Similarly, we can define a sampled-based approximation to the safe set as follows: $\tilde{\mathcal{SS}}^j_{\mathcal{G}} = \left\{ \bigcup_{k=0}^{\infty} \tilde{\mathcal{R}}^{\pi^j}_k(x_0^j) \ \bigcup \mathcal{G} \right\}$.

While $\tilde{\mathcal{SS}}^j_{\mathcal{G}}$ is not robust control invariant, with sufficiently many trajectory samples (i.e. $R$ sufficiently big), this approximation becomes more accurate in practice~\cite{SampleBasedLMPC}. To obtain a continuous approximation of the safe set for planning, we use the same technique as~\cite{SAVED}, and fit density model $\rho^{\mathcal{G}}_\alpha$ to $\bigcup_{k=0}^{j-1}\tilde{\mathcal{SS}}^{k}_{\mathcal{G}}$ and instead of enforcing the terminal constraint by checking if $x_{t+H} \in \bigcup_{k=0}^{j-1}\tilde{\mathcal{SS}}^{k}_{\mathcal{G}}$, \algabbr instead enforces that $\rho^{\mathcal{G}}_\alpha(x_{t+H}) > \delta$, where $\alpha$ is a kernel width parameter. We implement a tophat kernel density model using a nearest neighbors classifier with tuned kernel width $\alpha$ and use $\delta = 0$ for all experiments. Thus, all states within Euclidean distance $\alpha$ from the closest state in $\bigcup_{k=0}^{j-1}\tilde{\mathcal{SS}}^{k}_{\mathcal{G}}$ are considered safe under $\rho^{\mathcal{G}}_\alpha$.
\vspace{-0.15in}
\subsection{Start State Expansion Strategy}
\vspace{-0.01in}
\label{subsec:practical-expansion}
To provide a sample-based approximation to the procedure from Section~\ref{subsec:expansion}, we sample states $x_S^j$ from $\bigcup_{k=0}^j \tilde{\mathcal{SS}}^k$ and execute $\pi_{E, 0:H'-1}^j$ for $R$ trajectories of length $H'$, which approximate $\mathcal{M}$. We repeat this process until an $x_S^j$ is found such that all $R$ sampled trajectories satisfy the terminal state constraint that $x^j_{H'} \in \bigcup_{k=0}^j \tilde{\mathcal{SS}}^k_{\mathcal{G}}$ (Section~\ref{subsec:expansion}). Once such a state is found, a state is sampled from the last $H$ steps of the corresponding trajectories to serve as the start state for the next iteration, which approximates sampling from $\mathcal{M}_{H}$. We utilize a cost function which encourages controller domain expansion towards a specific desired start state $x^*$, although in general any cost function can be used. This cost function is interesting because it enables adaptation of a learning MPC controller to desired specifications while maintain controller feasibility. Precisely, we optimize a cost function which simply measures the discrepancy between a given state in a sampled trajectory and $x^*$, ie. $C_E^j(x, u) = D(x, x^*)$. This distance measure can be tuned on a task-specific basis based on the appropriate distance measures for the domain (Section~\ref{subsubsec:start-state-exp}). However, we remark that this technique requires: (1) an appropriate distance function $D(\cdot, \cdot)$ and (2) a reverse path from the goal to the start state, that may differ from the optimal forward path, along which the goal is robustly reachable.
\vspace{-0.15in}
\subsection{Goal Set Transfer}
\label{subsec:goal-set-selection}
We practically implement the goal set transfer strategy in Section~\ref{subsec:transfer} by fitting a new density model $\rho^\mathcal{H}_\alpha$ on the prefixes of prior trajectories that intersect some new user-specified goal set $\mathcal{H}$. If $\mathcal{H}$ is chosen such that $\tilde{\mathcal{SS}}^j_\mathcal{H}$ contains many states, the controller can seamlessly transfer to $\mathcal{H}$. If this is not the case, the controller domain for $\mathcal{H}$ must be expanded from $\mathcal{H}$ until it intersects many trajectories in the original domain.
\vspace{-0.15in}
\subsection{\algabbr Optimization Procedure}
As in prior work on MPC for nonlinear control~\cite{handful-of-trials, SAVED}, we solve the MPC optimization problem in~\eqref{eq:control-objective} over sampled open loop sequences of controls using the cross entropy method (CEM)~\cite{CEM}. In practice, we implement the terminal safe set constraints and state-space constraints in \eqref{eq:control-objective} and~\eqref{eq:reachability-policy-objective} by imposing a large cost on sampled action sequences which violate constraints when performing CEM. We use a probabilistic ensemble of $5$ neural networks to approximate $L^{\pi^j}_{\mathcal{G}}(x)$ as in~\cite{SAVED}. In contrast to~\cite{SAVED}, a separate $L^{\pi^j}_{\mathcal{G}}(x)$ is fit using data from each iteration instead of fitting a single function approximator on all data. We utilize Monte Carlo estimates of the cost-to-go values when fitting $L^{\pi^j}_{\mathcal{G}}(x)$. Each element of the ensemble outputs the parameters of a conditional axis-aligned Gaussian distribution and are trained on bootstrapped samples from the training dataset using a maximum likelihood~\cite{handful-of-trials}.
\vspace{-0.15in}

\section{Experiments}
\label{sec:experiments}
We evaluate whether \algabbr can enable (1) iterative improvement in expected performance for stochastic nonlinear systems, (2) adaptation to new start states and (3) transfer to new goal sets on $3$ simulated continuous control domains. In Section~\ref{subsec:domains} we describe the experimental domains, in Section~\ref{subsubsec:fixed-start-goal}, we evaluate the controller with fixed start states and goal sets, in Section~\ref{subsubsec:start-state-exp}, we expand the controller domain iteratively toward a desired start state far from the goal set, in Section~\ref{subsubsec:goal-transfer}, we switch the goal set during learning, and finally in Section~\ref{subsec:inverted_task} we utilize both start state expansion and the goal set transfer technique to control a pendulum to an upright position. In all experiments, we use $C(x, u) = \mathds{1}\{x \not\in \mathcal{G}\}$ as in \cite{SAVED}. Note that for this cost function, the maximum trajectory cost is the task horizon $T$, and the resulting objective corresponds to minimum time optimal control. We include comparisons to the minimum trajectory cost achieved by the state-of-the-art demonstration augmented model-based reinforcement learning algorithm, SAVED~\cite{SAVED} after $10$ iterations of training to evaluate the quality of the learned controller. For all experiments, we use $R=5$ closed-loop trajectories from the current controller to estimate $\tilde{\mathcal{SS}}^j_{\mathcal{G}}$ and perform start state expansion. Experimental domains have comparable stochasticity to those in~\cite{SampleBasedLMPC}. See Appendix~\ref{ssec:appendix-exp-details} for further details about experimental, optimization, and environment parameters. 
\vspace{-0.15in}
\subsection{Experimental Domains}
\label{subsec:domains}
\vspace{-0.25 in}
\begin{figure*}
\centering
\includegraphics[width=0.8\linewidth]{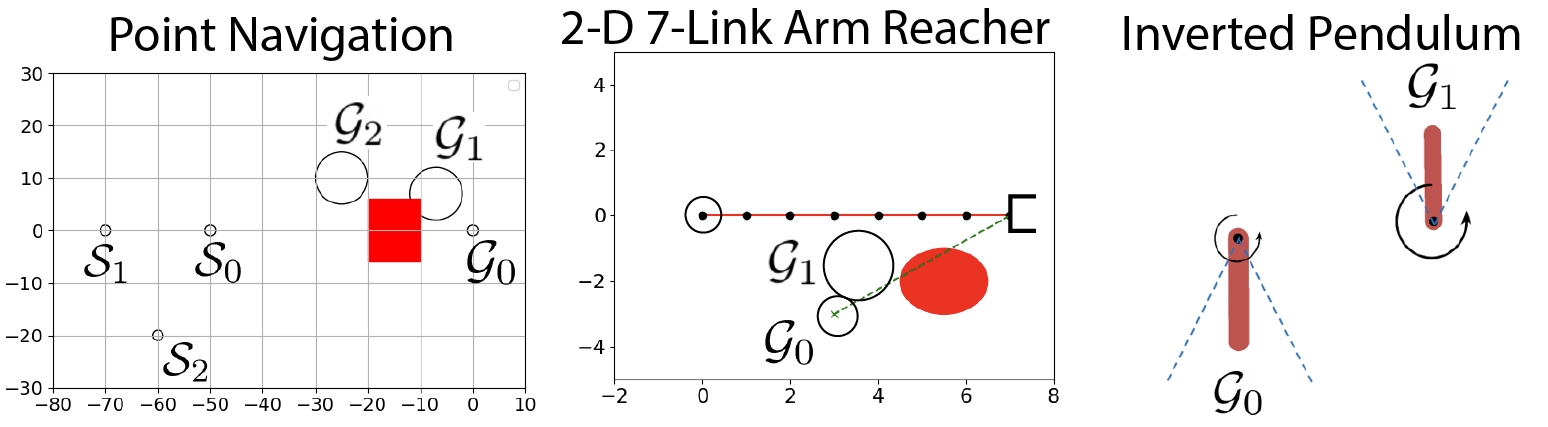}
\caption{\textbf{Experimental Domains: }We evaluate \algabbr on three stochastic domains: a navigation domain with an obstacle, a 2D 7-link arm reacher domain with an obstacle, and an inverted pendulum domain. In the first two domains, suboptimal demonstrations are provided, while no demonstrations are provided for the inverted pendulum task.}
\label{fig:domains}
\end{figure*}
\vspace{-0.2in}

\noindent\textbf{Point Mass Navigation: }We consider a 4-dimensional ($x$, $y$, $v_x$, $v_y$) navigation task as in~\cite{SAVED}, in which a point mass is navigating to a goal set (a unit ball centered at the origin unless otherwise specified). The agent exerts force $(f_x, f_y)$, $\lVert (f_x, f_y) \rVert \leq 1$, in each cardinal direction and experiences drag coefficient $\psi$. We introduce truncated Gaussian process noise $z_t\sim \mathcal{N}(0, \sigma^2 I)$ in the dynamics with domain $[-\sigma, \sigma]$. We include a large obstacle in the center of the environment that the robot must navigate around to reach the goal. While this task has linear dynamics, the algorithm must consider non-convex state space constraints and stochasticity.

\noindent\textbf{7-Link Arm Reacher: }Here, we consider a 2D kinematic chain with 7 joints where the agent provides commands in delta joint angles. We introduce truncated Gaussian process noise $z_t\sim \mathcal{N}(0, \sigma^2 I)$ in the dynamics with domain $[-\sigma, \sigma]$ and build on the implementation from~\cite{pyrobotics}. The goal is to control the end effector position to a $0.5$ radius circle in $\mathbb{R}^2$ centered at $(3, -3)$. We do not model self-collisions but include a circular obstacle of radius $1$ in the environment which the kinematic chain must avoid.

\noindent\textbf{Inverted Pendulum: }This environment is a noisy inverted pendulum task adapted from OpenAI Gym~\cite{openai-gym}. We introduce truncated Gaussian process noise in the dynamics.
\vspace{-0.2in}

\subsection{Fixed Start and Goal Conditions}
\label{subsubsec:fixed-start-goal}
\begin{figure*}
\centering
\begin{subfigure}[t]{.47\textwidth}
  \centering
  \includegraphics[width=0.6\linewidth]{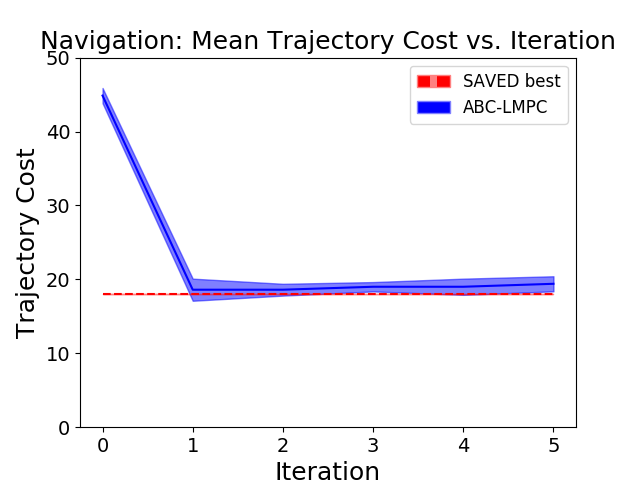}
\end{subfigure}
\hfill
\begin{subfigure}[t]{.47\textwidth}
  \centering
  \includegraphics[width=0.6\linewidth]{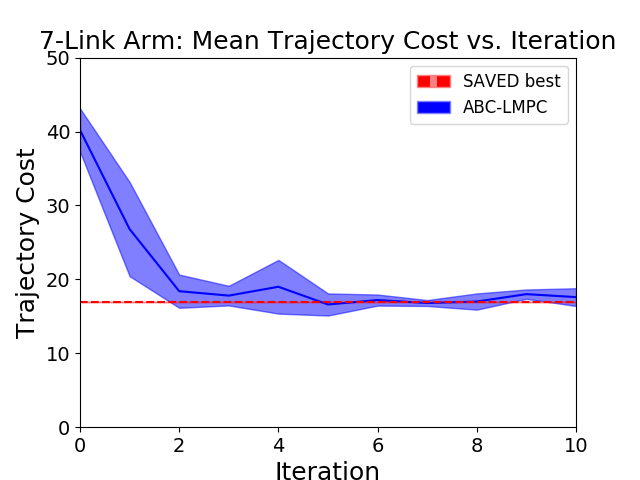}
\end{subfigure}
\caption{
\textbf{Fixed Start, Single Goal Set Experiments: } Learning curves for \algabbr averaged over $R=5$ rollouts per iteration on simulated continuous control domains when the start state and goal set is held fixed during learning. Performance of the demonstrations is shown at iteration 0, and the controller performance is shown thereafter. \textbf{Point Mass Navigation: } The controller immediately improves significantly upon the demonstration performance within 1 iteration, achieving a mean trajectory cost of around $20$ while demonstrations have mean trajectory cost of $42.58$.  \textbf{7-Link Arm Reacher: } The controller significantly improves upon the demonstrations, achieving a final trajectory cost of around $18$ while demonstrations achieve a mean trajectory cost of $37.77$. In all experiments, the controller quickly converges to the best cost produced by SAVED.}
\label{fig:single-start-exps}
\end{figure*}

We first evaluate \algabbr on the navigation and reacher environments with a fixed start state and goal set. In the navigation domain, the robot must navigate from $\mathcal{S}_0 = (-50, 0, 0, 0)$ to the origin ($\mathcal{G}_0$) while in the reacher domain, the agent must navigate from a joint configuration with the end effector at $(7, 0)$ to one with the end effector at $(3, -3)$ ($\mathcal{G}_1$). For optimization parameters and other experimental details, see Appendix~\ref{ssec:appendix-exp-details}. The controller rapidly and significantly improves upon demonstrations for both domains (Figure~\ref{fig:single-start-exps}). The controller achieves comparable cost to SAVED for both tasks and never violates constraints during learning.
\vspace{-0.3in}
\vspace{-0.55 in}
\subsection{Start State Expansion}
\label{subsubsec:start-state-exp}
\algabbr is now additionally provided a target start state which is initially outside its domain and learns to iteratively expand its domain toward the desired start state. We report the sequence of achieved start states over iterations in addition to the mean and standard deviation trajectory cost. \algabbr is able to maintain feasibility throughout learning and achieve comparable performance to SAVED at the final start state when SAVED is supplied with $100$ demonstrations from the desired state. To ensure that results are meaningful, we specifically pick desired start states such that given $100$ demonstrations from the original start state, \algabbr is never able to accomplish the task after 30 iterations of learning. A different $C_E(x, u)$ is used for start state expansion based on an appropriate distance metric for each domain.

We first consider a navigation task where $100$ suboptimal demonstrations are supplied from $(-25, 0, 0, 0)$ with average trajectory cost of $44.76$. The goal is to expand the controller domain in order to navigate from start states $\mathcal{S}_1 = (-70, 0, 0, 0)$ and $\mathcal{S}_2 = (-60, -20, 0, 0)$. $C_E(x, u)$ measures the Euclidean distance between the positions of $x$ and those of the desired start state. After $20$ iterations, the controller reaches the desired start state while consistently maintaining feasibility during learning (Table~\ref{start-state-exp}).

We then consider a similar Reacher task using the same suboptimal demonstrations from Section~\ref{subsubsec:fixed-start-goal}. The desired start end effector position is $(-1, 0)$, and $C_E(x, u)$ measures the Euclidean distance between the end effector position of states $x$ in optimized trajectories and that of the desired start state. Within $16$ iterations of learning, the controller is able to start at the desired start state while maintaining feasibility during learning (Table~\ref{start-state-exp}). On both domains, the controller achieves comparable performance to SAVED when trained with demonstrations from that start state and the controller successfully expands its domain while rapidly achieving good performance at the new states. Constraints are never violated during learning for all experiments.
\vspace{-0.35 in}
\begin{table}
\caption{\textbf{Start State Expansion Experiments: } \textbf{Pointmass Navigation: }Start State Expansion towards position $(-70, 0)$ (left) and $(-60, -20)$ (center). Here we see that \algabbr is able to reach the desired start state in both cases while consistently maintaining controller feasibility throughout learning. Furthermore, the controller achieves competitive performance with SAVED, which achieves a minimum trajectory cost of $21$ from $(-70, 0)$ and $23$ from $(-60, -20)$; \textbf{7-link Arm Reacher: }Here we expand the start state from that corresponding to an end effector position of $(7, 0)$ to that corresponding to an end effector position of $(-1, 0)$ (right). Again, we see that the controller consistently maintains feasibility during learning and achieves trajectory costs comparable to SAVED, which achieves a minimum trajectory cost of $24$.
The trajectory costs are presented in format: mean $\pm$ standard deviation over $R=5$ rollouts.}
\label{start-state-exp}
\begin{center}
\resizebox{\columnwidth}{!}{%
\begin{tabular}{|c|r|r|}
\multicolumn{3}{c}{\textbf{Point Navigation (-70, 0)}}\\ 
\toprule
Iteration & Start Pos (x, y) & Trajectory Cost\\
\midrule
4 &  $(-42.3, 1.33)$ & $23.0 \pm 0.89$\\ 
8 &  $(-54.1, 0.08)$ & $22.8 \pm 1.67$\\ 
12 & $(-61.2, 2.70)$ & $25.0 \pm 2.37$\\ 
16 & $(-70.3,-0.26)$ & $32.6 \pm 5.08$\\ 
20 & $(-70.4, 0.12)$ & $29.4 \pm 2.33$\\
\bottomrule
\end{tabular}
\quad
\begin{tabular}{|c|r|r|}
\multicolumn{3}{c}{\textbf{Point Navigation (-60, -20)}}\\ 
\toprule
Iteration & Start Pos (x, y) & Trajectory Cost\\
\midrule
4 &  $(-42.6, -8.76)$ & $19.6 \pm 4.22$\\ 
8 &  $(-54.6, -14.2)$ & $25.6 \pm 5.23$\\ 
12 & $(-58.8, -20.3)$ & $27.2 \pm 12.0$\\ 
16 & $(-60.6, -20.2)$ & $21.0 \pm 0.63$\\ 
20 & $(-60.5, -19.6)$ & $22.4 \pm 1.85$\\
\bottomrule
\end{tabular}
\quad
\begin{tabular}{|c|r|r|}
\multicolumn{3}{c}{\textbf{7-Link Reacher}}\\ 
\toprule
Iteration & Start EE Position & Trajectory Cost\\
\midrule
4 & $(-1.28, -0.309)$ & $31.6 \pm 8.04$\\ 
8 & $(-0.85, -0.067)$ & $30.8 \pm 15.7$\\ 
12 & $(-0.95, -0.014)$ & $20.2 \pm 1.83$\\ 
16 & $(-1.02, -0.023)$ & $19.4 \pm 4.03$\\ 
\bottomrule
\end{tabular}}
\end{center}
\end{table}
\vspace{-0.7 in}

\subsection{Goal Set Transfer}
\label{subsubsec:goal-transfer}
\begin{figure}
\begin{subfigure}{.33\textwidth}
  \centering
  \includegraphics[width=0.99\linewidth]{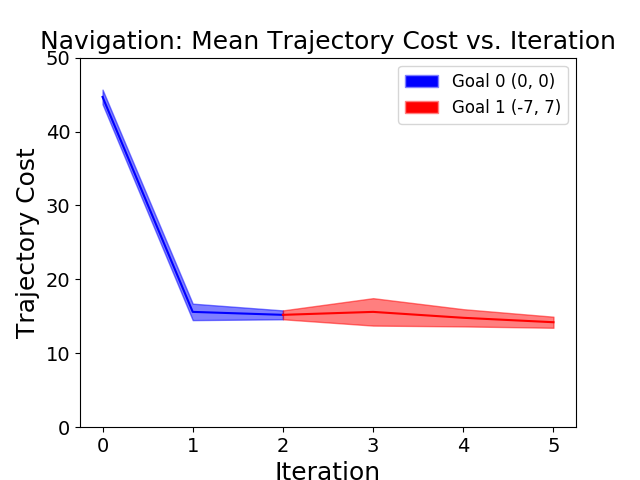}
\end{subfigure}%
\begin{subfigure}{.33\textwidth}
  \centering
  \includegraphics[width=0.99\linewidth]{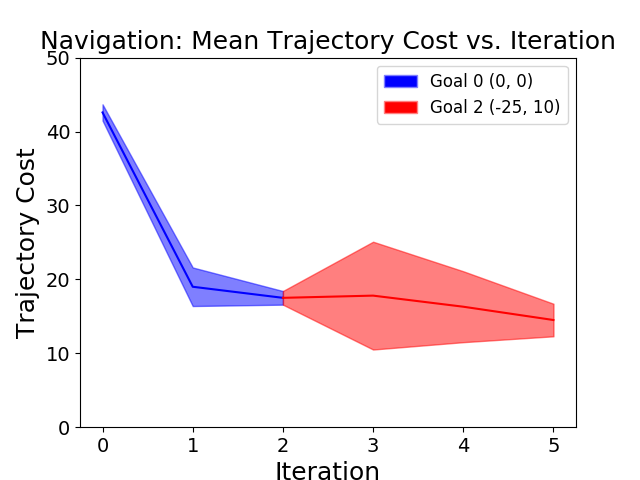}
\end{subfigure}
\begin{subfigure}{.33\textwidth}
  \centering
  \includegraphics[width=0.99\linewidth]{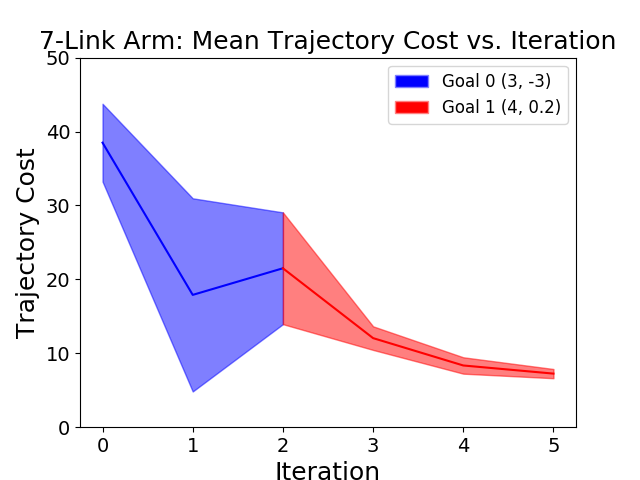}
\end{subfigure}
\caption{\textbf{Goal Set Transfer Learning:} In this experiment, the goal set is switched to to a new goal set at iteration $3$ and we show a learning curve which indicates performance on both the first goal (blue) and new goal (red). The controller is re-trained as in Section~\ref{subsec:goal-set-selection} to stabilize to the new goal. The controller immediately is able to perform the new task and never hits the obstacle. Results are plotted over $R=5$ controller rollouts per iteration.}
\label{fig:goal-set-transfer}
\end{figure}

\algabbr is trained as in Section~\ref{subsubsec:fixed-start-goal}, but after a few iterations, the goal set is changed to a new goal set that is in the controller domain. In the navigation domain, the robot is supplied a new goal set centered at $\mathcal{G}_1 = (-25, 10, 0, 0)$ or $\mathcal{G}_2 = (-7, 7, 0, 0)$ with radius $7$ after $2$ iterations of learning on the original goal set. We increase the radius so more prior trajectories can be reused for the new goal-conditioned value function. Results are shown in Figure~\ref{fig:goal-set-transfer} for both goal set transfer experiments. We also perform a goal set transfer experiment on the 7-link Reacher Task in which the robot is supplied a new goal set centered at $\mathcal{G}_1 = (4, 0.2)$ with radius $1$ after $2$ iterations of training. Results are shown in Figure~\ref{fig:goal-set-transfer}. In both domains, \algabbr seamlessly transfers to the new goal by leveraging prior experience to train a new set of value functions.
\vspace{-0.3 in}
\vspace{-0.4in}
\subsection{Inverted Pendulum Swing-Up Task}
\label{subsec:inverted_task}
In this experiment, we incorporate both the start state optimization procedure and goal set transfer strategies to balance a pendulum in the upright position, but without any demonstrations. We initialize the pendulum in the downward orientation ($\mathcal{G}_0$), and the goal of the task is to eventually stabilize the system to the upright orientation ($\mathcal{G}_1$). We iteratively expand the controller domain using the start state expansion strategy with initial goal $\mathcal{G}_0$ until the pendulum has swung up sufficiently close to the upright orientation. Once this is the case, we switch the goal set to $\mathcal{G}_1$ to stabilize to the upright position. The controller seamlessly transitions between the two goal sets, immediately transitioning to $\mathcal{G}_1$ while completing the task (convergence to either $\mathcal{G}_0$ or $\mathcal{G}_1$ within the task horizon) on all iterations (Table~\ref{pendulum-swing-up-exp}). $C_E(x, u)$ measures the distance between the pendulum's orientation and the desired start state's orientation.
\vspace{-0.3 in}
\begin{table}
\caption{\textbf{Pendulum Swing Up Experiment: }
We iteratively expand the controller domain outward from a goal set centered around the downward orientation ($\mathcal{G}_0$) towards the upward orientation until the controller domain includes a goal set centered around the upward orientation ($\mathcal{G}_1$). Then, the goal set is switched to $\mathcal{G}_1$. The resulting controller maintains feasibility throughout and seamlessly transitions to $\mathcal{G}_1$. The trajectory costs are presented as: mean $\pm$ standard deviation over $R=5$ rollouts. The upward orientation corresponds to a pendulum angle of $0^{\circ}$ and the angle (degrees) increases counterclockwise from this position until $360^{\circ}$.}
\vspace{0.1 in}
\label{pendulum-swing-up-exp}
	\begin{minipage}{0.5\linewidth}
		\centering
		\resizebox{0.8\columnwidth}{!}{%
        \begin{tabular}{|c|r|c|r|}
        \toprule
        Iteration & Start Angle & Goal Set & Trajectory Cost\\
        \midrule
        3 & $200.3$ & $\mathcal{G}_0$ & $30.2 \pm 1.47$\\ 
        6 & $74.3$ & $\mathcal{G}_0$ & $35.0 \pm 0.00$\\ 
        9 & $53.9$ & $\mathcal{G}_0$ & $34.4 \pm 0.49$\\ 
        12 & $328.1$ & $\mathcal{G}_1$ & $36.0 \pm 0.63$\\ 
        15 & $345.1$ & $\mathcal{G}_1$ & $13.8 \pm 7.03$\\ 
        18 & $0.6$ & $\mathcal{G}_1$ & $0.00 \pm 0.00$\\ 
        \bottomrule
        \end{tabular}}
	\end{minipage}\hfill
	\begin{minipage}{0.5\linewidth}
		\centering
		\includegraphics[width=0.38\linewidth]{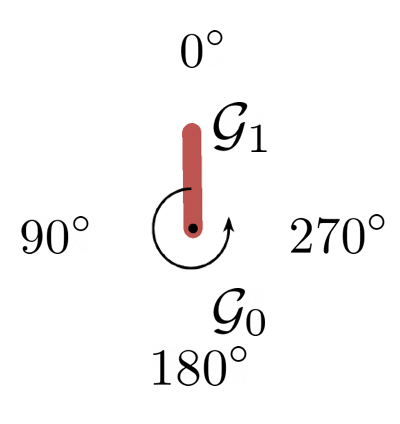}
	\end{minipage}
\end{table}
\vspace{-.4 in}
\section{Discussion and Future Work}
\label{sec:discussion}
We present a new algorithm for iteratively expanding the set of feasible start states and goal sets for an LMPC-based controller and provide theoretical guarantees on iterative improvement in expectation for non-linear systems under certain conditions on the cost function and demonstrate its performance on stochastic linear and nonlinear continuous control tasks. In future work, we will explore synergies with sample based motion planning to efficiently generate asymptotically optimal plans. We will also integrate the reachability-based domain expansion strategies of \algabbr with model-based RL to learn safe and efficient controllers when dynamics are learned from experience.
\\
\\
\noindent \textbf{Acknowledgements:} \begin{scriptsize}
\noindent This research was performed at the AUTOLAB at UC Berkeley in affiliation with the Berkeley AI Research (BAIR) Lab. Authors were also supported by the Scalable Collaborative Human-Robot Learning (SCHooL) Project, a NSF National Robotics Initiative Award 1734633, and in part by donations from Google and Toyota Research Institute. Ashwin Balakrishna is supported by an NSF GRFP. This article solely reflects the opinions and conclusions of its authors and does not reflect the views of the sponsors. We thank our colleagues who provided helpful feedback and suggestions, especially Michael Danielczuk, Daniel Brown and Suraj Nair.
\end{scriptsize}
\vspace{-0.1in}
\renewcommand*{\bibfont}{\footnotesize}
\printbibliography %
\clearpage
\appendix
\section{Appendix}
\label{sec:appendix}

\subsection{Proofs of Controller Properties}
\label{ssec:appendix-proofs}

\paragraph{Proof of \Cref{lemma:recursive_feasibility}}
We proceed by induction. By assumption~\ref{ass:safe_set_initial_condition}, $J^0_{0 \rightarrow H}(x_0^j) < \infty$. By the definition of $V_{\mathcal{G}}^{\pi^j}$ and $\mathcal{F}_{\mathcal{G}}^j$, $J^j_{0 \rightarrow H}(x_0^j) < \infty$. Let $J^j_{t \rightarrow t+H}(x_t^j) < \infty$ for some $t\in \mathbb{N}$. In the following expressions, we do not explicitly write the MPC problem constraints for clarity. Conditioning on the random variable $x^j_t$:

\begin{align}
    J^j_{t \rightarrow t+H}(x_t^j) &= \underset{w^j_{t:t+H-1}}{\mathds{E}}\left[\sum_{k=0}^{H-1} C(x^j_{t+k|t}, \pi^{*,j}_{t+k|t}(x^j_{t+k|t})) + V_{\mathcal{G}}^{\pi^{j-1}}(x^j_{t+H|t})\right]\label{P1.1}\\
    &= C(x_t^j, \pi_{t|t}^{*,j}(x_t^j)) + \mathds{E}_{w^j_{t:t+H-1}}\left[\sum_{k=1}^{H-1} C(x^j_{t+k|t}, \pi^{*,j}_{t+k|t}(x^j_{t+k|t})) + V_{\mathcal{G}}^{\pi^{j-1}}(x^j_{t+H|t})\right]\label{P1.2}\\
    \begin{split}
    &= C(x_t^j, \pi_{t|t}^{*,j}(x_t^j))\\
    &+ \underset{w^j_{t:t+H}}{\mathds{E}}\bigg[\sum_{k=1}^{H-1} C(x^j_{t+k|t}, \pi^{*,j}_{t+k|t}(x^j_{t+k|t})) + C(x^j_{t+H|t}, \pi^{l}(x^j_{t+H|t}))\\
    &+ V_{\mathcal{G}}^{\pi^{j-1}}(x^j_{t+H+1|t})\bigg], \ l \in [j-1]\label{P1.3}\\
    \end{split}\\
    \begin{split}
    \\
    &\geq C(x_t^j, \pi_{t|t}^{*,j}(x_t^j))\\
    &+ \underset{w^j_t}{\mathds{E}}\bigg[\min_{\pi_{t+1:t+H|t+1}} \underset{w^j_{t+1:t+H}}{\mathds{E}}\bigg[\sum_{k=1}^{H-1} C(x^j_{t+k|t+1}, \pi_{t+k|t+1}(x^j_{t+k|t+1})) \\
    &+ C(x^j_{t+H|t+1}, \pi_{t+H|t+1}(x^j_{t+H|t+1})) \\ &+ V_{\mathcal{G}}^{\pi^{j-1}}(x^j_{t+H+1|t+1})\bigg]\bigg]\\\label{P1.4}
    \end{split}
    \\&= C(x_t^j, \pi^{j}(x_t^j)) + \underset{w_t^j}{\mathds{E}}\left[J^j_{t+1 \rightarrow t + H + 1}(x_{t + 1}^j)|x_t^j\right]\label{P1.5}
    \end{align}
Equation~\ref{P1.1} follows from the definition in \ref{eq:control-objective}, equation~\ref{P1.3} follows from the definition of $V_{\mathcal{G}}^{\pi_{j-1}}$, which is defined as a point-wise minimum over $\left(L^{\pi^l}_\mathcal{G}\right)_{l=0}^{j-1}$. We take a function $L^{\pi^l}_\mathcal{G}$ that is active at $x_{t+H|t}^j$ and apply its definition to expand it and then replace $L^{\pi^l}_\mathcal{G}$ with $V^{\pi^{j-1}}_\mathcal{G}$ in the expansion. The inner expectation in equation~\ref{P1.4} conditions on the random variable $x^j_{t+1}$, and the outer expectation integrates it out. The inequality in~\ref{P1.4} follows from the fact that $[\pi^{*,j}_{t+1|t},\ldots,\pi_{t+H-1|t}^{*,j}, \pi^{j-1}]$ is a possible solution to~\eqref{P1.4}. Equation~\ref{P1.5} follows from the definition in equation~\ref{eq:control-objective}.\\
We have shown that $J^j_{t \rightarrow t+H}(x_t^j)<\infty \implies \underset{w_{t}^j}{\mathds{E}}\left[J^j_{t+1 \rightarrow t + H + 1}(x_{t + 1|t}^j)\right] < \infty$. So:
\begin{align}
    \underset{w_{0:t-1}^j}{\mathds{E}}\left[J^j_{t \rightarrow t+H}(x_t^j)\right]<\infty \implies & \underset{w_{0:t-1}^j}{\mathds{E}}\left[\underset{w_{t}^j}{\mathds{E}}\left[J^j_{t+1 \rightarrow t + H + 1}(x_{t + 1|t}^j)\right]\right] \\
    = & \underset{w_{0:t}^j}{\mathds{E}}\left[J^j_{t+1 \rightarrow t + H + 1}(x_{t + 1}^j)\right] < \infty
\end{align}
By induction, $\underset{w^j_{0:t-1}}{\mathds{E}}[J^j_{t\rightarrow t + H}(x_t^j)] < \infty$ $\forall t\in \mathbb{N}$. Therefore, the controller is feasible at iteration $j$. \qed

\paragraph{Proof of \Cref{lemma:convergence}}
By \Cref{lemma:recursive_feasibility} and Assumption~\ref{ass:cost}, $\forall L \in \mathbb{N}$,
\begin{align}
    &\underset{w^j_{1:L-1}}{\mathds{E}}\left[\sum_{k=0}^{L-1} C(x_k^j, \pi^j(x_k^j)) + J^j_{L\rightarrow L+H} (x_L^j)\right] \leq J_{0 \rightarrow H}^j(x_0^j)\label{L1.1}\\
    &\implies\ \underset{w_{1:L-1}^j}{\mathds{E}}\left[J^j_{L\rightarrow L+H} (x_L^j)\right] \leq J_{0 \rightarrow H}^j(x_0^j) - \underset{w^j_{1:L-1}}{\mathds{E}}\left[\sum_{k=0}^{L-1} C(x_k^j, \pi^j(x_k^j))\right]\\ &\leq J_{0 \rightarrow H}^j(x_0^j) - \epsilon\sum_{k=0}^{L-1}P(x^j_k \not\in\mathcal{G})\label{L1.2}
\end{align}
Line~\ref{L1.2} follows from rearranging~\ref{L1.1} and applying assumption~\ref{ass:cost}. Because $\mathcal{G}$ is robust control invariant by assumption~\ref{ass:robust_control_invariant_goal}, $x_t \in \mathcal{G} \implies x_{t+k} \in \mathcal{G}\ \forall k \geq 0$. Now, assume  $\lim_{k \rightarrow \infty} P(x_k^j \not\in\mathcal{G})$ does not exist or is nonzero. This implies that $P(x_k^j \not\in\mathcal{G}) \geq \delta > 0$ infinitely many times. By the Archimedean principle, the RHS of~\ref{L1.2} can be driven arbitrarily negative, which is impossible. By contradiction, $\lim_{k \rightarrow \infty} P(x_k^j \not\in\mathcal{G}) = 0$.\qed

\paragraph{Proof of \Cref{theorem:improvement}}

Let $j \in \mathbb{N}$
\begin{align}
    J_{0\rightarrow H}^{j}(x_0) &\geq C(x_0, u_0) + \underset{w_{0}^j}{\mathds{E}}\left[J^j_{1\rightarrow H+1}(x^j_{1})\right]\label{P2.1}\\
    &\geq \underset{w^j}{\mathds{E}}\left[\sum_{t=0}^\infty C(x^j_t, \pi^j(x^j_t))\right] + \lim_{t\rightarrow \infty} \underset{w_{0:t-1}^j}{\mathds{E}}\left[J^j_{t\rightarrow t+H}(x_t^j)\right]\label{P2.2}\\
    &= \underset{w^j}{\mathds{E}}\left[\sum_{t=0}^\infty C(x^j_t, \pi^j(x^j_t))\right] + \lim_{t\rightarrow \infty} \underset{\mathds{1}\{x_t^j \not\in\mathcal{G}\}}{\mathds{E}}\left[ \underset{w_{0:t-1}^j}{\mathds{E}}\left[J^j_{t\rightarrow t+H}(x_t)|\mathds{1}\{x_t^j \not\in\mathcal{G}\}\right]\right]\label{P2.3}\\
    &= \underset{w^j}{\mathds{E}}\left[\sum_{t=0}^\infty C(x^j_t, \pi^j(x^j_t))\right] + \lim_{t\rightarrow \infty} \underset{w_{0:t-1}^j}{\mathds{E}}\left[J^j_{t\rightarrow t+H}(x^j_t)|x^j_t \not\in\mathcal{G}\right]P(x_t^j \not\in\mathcal{G})\label{P2.4}
\end{align}
\begin{align}
    &\geq \underset{w^j}{\mathds{E}}\left[\sum_{t=0}^\infty C(x^j_t, \pi^j(x^j_t))\right] + \lim_{t\rightarrow \infty} \epsilon P(x_t^j \not\in\mathcal{G})\label{P2.4a}\\
    &= \underset{w^j}{\mathds{E}}\left[\sum_{t=0}^\infty C(x_t^{j}, \pi^j(x^{j}_t))\right] = J^{\pi^j} (x_0)\label{P2.5}
\end{align}
Equations~\ref{P2.1} and~\ref{P2.2} follow from repeated application of \Cref{lemma:recursive_feasibility}~\eqref{P1.5}. Equation~\ref{P2.3} follows from iterated expectation, equation~\ref{P2.4} follows from the cost function assumption~\ref{ass:cost}. Equation~\ref{P2.4a} follows again from assumption~\ref{ass:cost} (incur a cost of at least $\epsilon$ for not being at the goal at time $t$). Then,  Equation \ref{P2.5} follows from \Cref{lemma:convergence}.
Using the above inequality with the definition of $J^{\pi^j}(x_0)$,
\begin{align}
    J^{j}_{0\rightarrow H}(x_0) \geq J^{\pi^j}(x_0) &= \underset{w_{0:H-1}^j}{\mathds{E}}\left[\sum_{t=0}^{H-1} C(x^j_t, \pi^j(x_t)) + V_{\mathcal{G}}^{\pi^j}(x^j_H)\right]\label{P2.6}\\
    &\geq \underset{w_{0:H-1}^j}{\mathds{E}}\left[ \sum_{t=0}^{H - 1} C(x^j_t, \pi^{*,j}_{t|0}(x_{t|0})) + V_{\mathcal{G}}^{\pi^{j}}(x_{H|0})\right] = J^{j+1}_{0\rightarrow H}(x_0)\label{P2.7}\\
    &\geq J^{\pi^{j+1}}(x_0)\label{P2.8}
\end{align}
Equation~\ref{P2.6} follows from equation~\ref{P2.5}, equation~\ref{P2.7} follows from taking the minimum over all possible $H$-length sequences of policies in the policy class $\Pi$. Equation~\ref{P2.8} follows from equation~\ref{P2.5}. By induction, this proves the theorem.

Note that this also implies convergence of $(J^{\pi^j}(x_0))_{j=0}^\infty$ by the Monotone Convergence Theorem. \qed

\paragraph{Proof of \Cref{lemma:expansion}}

The proof is identical to~\cite{SampleBasedLMPC}. Because $\mathcal{SS}^j_{\mathcal{G}}$ is an increasing sequence of sets, $\mathcal{F}^j_\mathcal{G}$ is also an increasing sequence of sets by definition.\qed

\subsection{\algnameonly Implementation Details}
\label{ssec:appendix-implementation}

\subsubsection{Solving the MPC Problem}
As in~\cite{SAVED}, we sample a fixed population size of action sequences at each iteration of CEM from a truncated Gaussian. These action sequences are simulated over a known model of the system dynamics and then the sampling distribution for the next iteration is updated based on the lowest cost sampled trajectories. For the cross entropy method we build off of the implementation in~\cite{handful-of-trials}. Precisely, at each timestep in a trajectory, a conditional Gaussian is initialized with the mean based on the final solution for the previous timestep and some fixed variance. Then, at each iteration of CEM, pop\_size action sequences of plan\_hor length are sampled from the conditional Gaussian, simulated over a model of the system dynamics, and then the num\_elites samples with the lowest sum cost are used to refit the mean and variance of the conditional Gaussian distribution for the next iteration of CEM. This process is repeated num\_iters times. The sum cost of an action sequence is computed by summing up the task cost function at each transition in the resulting simulated trajectory and then adding a large penalty for each constraint violating state in the simulated trajectory and an additional penalty if the terminal state in the simulated trajectory does not have sufficient density under $\rho_{\mathcal{G}}$. For all experiments, we add a $1e6$ penalty for violating terminal state constraints and a $1e8$ penalty for violating task constraints. In practice to accelerate domain expansion to $x^*$, when selecting initial states $x_S^j$ from $\bigcup_{k=0}^j \tilde{\mathcal{SS}}^k$, we sort states in the safeset under $C_E^j(x)$ and use this to choose $x_S^j$ close to $x^*$ under $C_E^j(x)$. Note that this choice does not impact any of the theoretical guarantees.

\subsubsection{Value Function}
We represent each member of the probabilistic ensemble of neural networks used to approximate $L^{\pi^j}(x)$ with a neural network with 3 hidden layers, each with 500 hidden units. We use swish activations, and update weights using the Adam Optimizer with learning rate $0.001$. We use 10 epochs to learn the weights for $L^{\pi^j}(x)$.

\subsubsection{Start State Expansion}
We again perform trajectory optimization using the cross entropy method and for each experiment use the same pop\_size, num\_elites, num\_iters parameters as for solving the MPC problem. Costs for action sequences are computed by summing up $C^j_E(x)$ evaluated at each state $x$ in the corresponding simulated trajectory, and the same mechanism is used for enforcing the terminal state constraint and task constraints as for solving the MPC problem.

\subsection{Experiment Specific Parameters}
\label{ssec:appendix-exp-details}

\subsubsection{Pointmass Navigation}
\paragraph{Environment Details: }
 We use $\psi = 0.2$ and $\sigma = 0.05$ in all experiments in this domain. Demonstration trajectories are generated by guiding the robot past the obstacle along a very suboptimal hand-tuned trajectory for the first half of the trajectory before running LQR with clipped actions on a quadratic approximation of the true cost. Gaussian noise is added to the demonstrator controller. The task horizon is set to $T = 50$.
\paragraph{Task Controller MPC Parameters: }  For the single start, single goal set case we use popsize = 400, num\_elites = 40, cem\_iters = 5, and plan\_hor = 15. For all start state expansion experiments, we utilize the same popsize, num\_elites, and cem\_iters but utilize plan\_hor = 20. For experiments we utilize $\alpha = 2$ for the kernel width parameter for density model $\rho^{\mathcal{G}}_\alpha$.
\paragraph{Start State Expansion Parameters: } We utilize $H' = H - 5$ for all experiments (trajectory optimization horizon for exploration policy).
\paragraph{SAVED Baseline Experimental Parameters: } We supply SAVED with $100$ demonstrations generated by the same demonstration policy as for \algabbr. We utilize $\alpha = 3$ and utilize the implementation from~\cite{SAVED}. Both the value function and dynamics for SAVED are represented with a probabilistic ensemble of $5$ neural networks with 3 hidden layers of 500 hidden units each. We use swish activations, and update weights using the Adam Optimizer with learning rate $0.001$.

\subsubsection{7-Link Reacher Arm}
\paragraph{Environment Details: }
We use $\sigma = 0.03$ for all experiments. The state space consists of the 7 joint angles. Each link is of 1 unit in length and the goal is to control the end effector position to a $0.5$ radius circle in $\mathbb{R}^2$ centered at $(3, -3)$. We do not model self-collisions but also include a circular obstacle of radius $1$ in the environment which the kinematic chain must navigate around. Collisions with the obstacle are checked by computing the minimum distance between each link in the kinematic chain and the center of the circular obstacle and determining whether any link has a minimum distance from the center of the obstacle that is less than the radius of the obstacle. The task horizon is set to $T = 50$. We build on the implementation provided through~\cite{pyrobotics}.
\paragraph{Task Controller MPC Parameters: }  For the single start, single goal set case we use popsize = 400, num\_elites = 40, cem\_iters = 5, and plan\_hor = 15. For all start state expansion experiments, we utilize the same popsize, num\_elites, and cem\_iters but utilize plan\_hor = 20. For experiments we utilize $\alpha = 0.5$ for the kernel width parameter for density model $\rho^{\mathcal{G}}_\alpha$.
\paragraph{Start State Expansion Parameters: } We utilize $H' = H - 5$ for all experiments (trajectory optimization horizon for exploration policy).
\paragraph{SAVED Baseline Experimental Parameters: }  We supply SAVED with $100$ demonstrations generated by the same demonstration policy as for \algabbr. We utilize $\alpha = 0.5$ and utilize the implementation from~\cite{SAVED}. Both the value function and dynamics for SAVED are represented with a probabilistic ensemble of $5$ neural networks with 3 hidden layers of 500 hidden units each. We use swish activations, and update weights using the Adam Optimizer with learning rate $0.001$.

\subsubsection{Inverted Pendulum}
\paragraph{Environment Details: }
 We use $\sigma = 0.5$ for all experiments.
 The robot consists of a single link and can exert a torque to rotate it.  The state space consists of the angle and angular velocity of the pendulum. Note that there are only stable orientations, the upright orientation and downward orientation for this task, and thus for a goal set to be robust control invariant, it will likely need to be defined around the neighborhood of these orientations. The task horizon is set to $T = 40$. We define $\mathcal{G}_1$ as the goal set centered around the downward orientation and $\mathcal{G}_2$ as the goal set centered around the upright orientation. Precisely, inclusion in $\mathcal{G}_1$ is determined by determining whether the orientation of the pendulum is within 45 degrees of the downward orientation. Similarly, inclusion in $\mathcal{G}_2$ is determined by determining whether the orientation of the pendulum is within 45 degrees of the upward orientation.
\paragraph{Task Controller MPC Parameters: }  We utilize popsize = 600, num\_elites = 40, cem\_iters = 5, and plan\_hor = 15. For experiments we utilize $\alpha = 2$ for the kernel width parameter for density model $\rho^{\mathcal{G}}_\alpha$. 
\paragraph{Start State Expansion Parameters: } We utilize $H' = H$ for all experiments (trajectory optimization horizon for exploration policy).

\subsection{Controller Domain Expansion Strategy}
\label{appendix:controller_spec}

Here we discuss how the controller domain can be  expanded when the safe set and value function are updated based on samples from the exploration policy. To approximately expand $\bigcup_{k=0}^j \mathcal{SS}^k_{\mathcal{G}}$, we can again solve the following 1-step trajectory optimization problem:

\begin{small}
\begin{align}
\begin{split}
 \pi_{E, 0:H'-1}^j = \myargmin{\pi_{0:H'-1}\in \Pi^{H'}} \quad &\underset{w^j_{0:H'-2}}{\mathds{E}}\left[\sum_{i=0}^{H'-1} C_E^j(x^j_{i}, \pi_{i}(x^j_{i}))\right]\\
\text{s.t. } \quad & x^j_{i+1} = f(x^j_{i}, \pi_{i}(x^j_{i}), w_i) \ \forall i \in \{0,\ldots,H'-1\}\\
& x^j_{H'} \in \bigcup_{k=0}^{j-1} \mathcal{SS}^k_{\mathcal{G}},\ \forall w_{0:H'-2} \in \mathcal{W}^{H'-1}\\
& x^j_{0:H'} \in \mathcal{X}^{H'+1},\ \forall w_{0:H'-2} \in \mathcal{W}^{H'-1}
\end{split}
\end{align}
\end{small}

For all $x_S^j \in \mathcal{SS}^{j-1}_{\mathcal{G}}$, the states $\bigcup_{k=0}^{H'} \mathcal{R}_k^{\pi_{E,0:H'-1}^j}(x^j_S) \cup \bigcup_{k=1}^\infty \mathcal{R}_k^{\pi^j}(\mathcal{R}_{H'}^{\pi_{E,0:H'-1}^j}(x^j_S))$ are added to $\mathcal{SS}^{j}_{\mathcal{G}}$. The second union is included to define the value function for the composition of $\pi^j$ and $\pi_{E,0:H'-1}^j$. This is analogous to running the exploration policy followed by running the task-directed policy $\pi^j$.
Denoting the safe set where $\pi^j$ is executed as $\mathcal{SS}^{\pi^j}_{\mathcal{G}} = \bigcup_{k=1}^\infty \mathcal{R}_k^{\pi^j}(\mathcal{R}_{H'}^{\pi_{E,0:H'-1}^j}(\mathcal{SS}^{j-1}_{\mathcal{G}})) \cup \bigcup_{k=1}^\infty \mathcal{R}_k^{\pi^j}(\mathcal{SS}^{j-1}_{\mathcal{G}})$, we redefine $L_\mathcal{G}^{\pi^j}$ as:
\begin{align}
     L^{\pi^{j}}_\mathcal{G}(x) = \begin{cases} 
          \underset{w}{\mathds{E}}\left[C(x, \pi^j(x)) + L_{\mathcal{G}}^{\pi^j}(f(x, \pi^j(x), w))\right] & x\in \mathcal{SS}^{\pi^j}_{\mathcal{G}} \setminus \mathcal{G} \\
          \underset{w}{\mathds{E}}\left[C(x, \pi^j_{E,0:H'-1}(x)) + L_{\mathcal{G}}^{\pi^j}(f(x, \pi^j_{E,0:H'-1}(x), w))\right] & x\in \mathcal{SS}^{j}_{\mathcal{G}} \setminus \mathcal{SS}^{\pi^j}_{\mathcal{G}} \\
          0 & x \in \mathcal{G}\\
          +\infty & x \not\in\mathcal{SS}^j_{\mathcal{G}}
      \end{cases}
\end{align}
This means that trajectories from the exploration policy can spend more time outside of the safe set. In either case, the safe set remains robust control invariant.

Thus, each iteration $j$ is split into two phases. In the first phase, $\pi^j$ is executed and in the second phase, $\pi^j_{E, 0:H'-1}$ is executed. This procedure provides a simple algorithm to expand the policy's domain $\mathcal{F}_{\mathcal{G}}^j$ while still maintaining its theoretical properties.

\end{document}